\documentclass[conference]{IEEEtran}
\usepackage{amsmath, amssymb, bm, cite, epsfig, psfrag}
\usepackage{graphicx}
\usepackage{soul} 
\usepackage[margin=0.625in]{geometry}
\usepackage{dblfloatfix}
\usepackage{array}
\usepackage{lipsum}
\usepackage{subcaption}

\newcolumntype{P}[1]{>{\centering\hspace{0pt}}p{#1}}
\newcolumntype{M}[1]{>{\centering\hspace{0pt}}m{#1}}
\newcolumntype{L}{>{\centering\arraybackslash}m{3cm}}
\usepackage[font=small]{caption}
\usepackage{epstopdf}	
\usepackage{longtable}
\usepackage{supertabular,booktabs}
\usepackage{bbm}
\usepackage{multirow}
\usepackage[usenames,dvipsnames]{xcolor}

\usepackage{etoolbox}
\usepackage{pbox}
\usepackage{fixltx2e}
\usepackage{tabu}	
\usepackage{filecontents}
\usepackage{enumerate}
\usepackage{textcomp}
\usepackage{makecell}
\usepackage{gensymb}
\usepackage{colortbl}
\usepackage{fancyhdr}

\usepackage{bm}

\newtoggle{conference}

\togglefalse{conference} 
\graphicspath{{figures/}}

\newcolumntype{?}{!{\vrule width 2pt}}

\fancypagestyle{firststyle}{
	\fancyhf{}
	\fancyhead[L]{S. Ju and T. S. Rappaport, ``140 GHz Urban Microcell Propagation Measurements for Spatial Consistency Modeling,''  2021 \textit{IEEE International Conference on Communications (ICC)}, Jun. 2021, pp. 1-6.}     

}
\IEEEoverridecommandlockouts

\begin{document}
\title{140 GHz Urban Microcell Propagation Measurements for Spatial Consistency Modeling} 
\author{\IEEEauthorblockN{Shihao Ju and Theodore S. Rappaport}

\IEEEauthorblockA{	\small NYU WIRELESS, Tandon School of Engineering, New York University, Brooklyn, NY, 11201\\
				\{shao, tsr\}@nyu.edu}
					\thanks{This research is supported by the NYU WIRELESS Industrial Affiliates Program and National Science Foundation (NSF) Research Grants: 1909206 and 2037845.}
}

\maketitle
\thispagestyle{firststyle}
\begin{abstract}
Sub-Terahertz frequencies (frequencies above 100 GHz) have the potential to satisfy the unprecedented demand on data rate on the order of hundreds of Gbps for sixth-generation (6G) wireless communications and beyond. Accurate beam tracking and rapid beam selection are increasingly important since antenna arrays with more elements generate narrower beams to compensate for additional path loss within the first meter of propagation distance at sub-THz frequencies. Realistic channel models for above 100 GHz are needed, and should include spatial consistency to model the spatial and temporal channel evolution along the user trajectory. This paper introduces recent outdoor urban microcell (UMi) propagation measurements at 142 GHz along a 39 m $\times$ 12 m rectangular route (102 m long), where each consecutive and  adjacent receiver location is 3 m apart from each other. The measured power delay profiles and angular power spectrum at each receiver location are used to study spatial autocorrelation properties of various channel parameters such as shadow fading, delay spread, and angular spread along the track. Compared to the correlation distances reported in the 3GPP TR 38.901 for frequencies below 100 GHz, the measured correlation distance of shadow fading at 142 GHz (3.8 m) is much shorter than the 10-13 m as specified in 3GPP; the measured correlation distances of delay spread and angular spread at 142 GHz (both 12 m) are comparable to the 7-10 m as specified in 3GPP. This result may guide the development of a statistical spatially consistent channel model for frequencies above 100 GHz in the UMi street canyon environment. 

\end{abstract}
    
\begin{IEEEkeywords}                            
Terahertz; Spatial Consistency; Channel Measurement; Channel Modeling; 140 GHz; 142 GHz; 5G; 6G 
\end{IEEEkeywords}

\section{Introduction}
Emerging applications such as wireless cognition and sensing accelerate wireless communication research at THz frequencies (100 GHz - 3 THz). As the fifth-generation (5G) wireless networks start operating at frequencies above 24 GHz, the spectrum above 100 GHz is being considered as a key feature of the sixth-generation (6G) wireless communication technologies due to vastly unused bandwidth of many tens of GHz \cite{Rap19access}.

One critical challenge of wireless communications above 100 GHz is severe path loss in the first meter of propagation from the transmitting antenna and signal attenuation through partitions (e.g., 4-8 dB higher loss at 142 GHz than 28 GHz for different materials \cite{Xing19globecom}); thus, antenna arrays with massive antenna elements having very narrow beamwidth and very high beamforming gain will be required to compensate for the additional path loss in the first meter of propagation and the larger partition losses at sub-THz frequencies \cite{Xing21a,Xing21b,Ju21jsac}. On the contrary, reflecting objects such as walls and lampposts have more energy in the reflected direction above 100 GHz \cite{Xing21a,Xing21b,Xing21icc}. Narrow beams require rapid beam steering and accurate beam tracking to support multiple moving user terminals (UT) since highly directional channels are sensitive to random blockages by humans or vehicles in the environment \cite{Kanhere21vtc,Giordani19CST}. To evaluate different beam tracking schemes, an accurate channel model for frequencies above 100 GHz is required, and should be able to generate time-variant channel impulse responses according to UT mobility \cite{Xing21icc}. Therefore, as a vital modeling component, spatial consistency was proposed in the third generation partnership project (3GPP) channel model and incorporated in other channel models such as NYUSIM, COST 2100, METIS, and mmMAGIC \cite{Ju19gc,3GPP38901r16,METIS15,mmMAGIC17}. 

Spatial consistency represents the phenomenon that a moving UT or multiple closely-located UTs experience a similar scattering environment in a local area (e.g., within 15 m), indicating the channels across these locations are spatially correlated \cite{Ju18gc}. To emulate this local spatial correlation, the conventional statistical channel model must be able to generate channel impulse responses with spatially consistent large-scale parameters such as shadow fading, line-of-sight/non-line-of-sight (LOS/NLOS) condition, and small-scale parameters such as the delay, power, and angles of each cluster and multipath component \cite{Ju19gc}.

The spatial autocorrelation properties of shadow fading and delay spread over distance at sub-6 GHz and mmWave frequencies have been well studied \cite{Zhang08gc,Guan15a,Rap17tap,Dai20WSA}, but similar studies have not been conducted at frequencies above 100 GHz. This paper investigates the spatial autocorrelation property for frequencies above 100 GHz based on the 142 GHz outdoor urban microcell (UMi) measurements conducted along a rectangular route. The spatial autocorrelation functions of shadow fading, root mean square (RMS) delay spread, and angular spread are derived, and the corresponding correlation distances for sub-THz frequencies are proposed.

The remainder of this paper is organized as follows. Section II introduces the 142 GHz measurement system, environment, and procedure. Section III presents the large-scale path loss and shadow fading along the rectangular route. Section IV derives the spatial autocorrelation functions of shadow fading, delay spread, and angular spread. Finally, concluding remarks in Section V show that in the UMi street canyon scenario, the correlation distance of shadow fading at 142 GHz (3.8 m) is much shorter than 10-13 m for frequencies below 100 GHz while the correlation distances of delay spread and angular spread at 142 GHz (both 12 m) are comparable to 7-10 m for frequencies below 100 GHz as reported in 3GPP TR 38.901 \cite{3GPP38901r16}.


\section{142 GHz Outdoor Local Area Measurement Campaign}
\begin{figure*}[h!]
	\centering
	\includegraphics[width=0.8\linewidth]{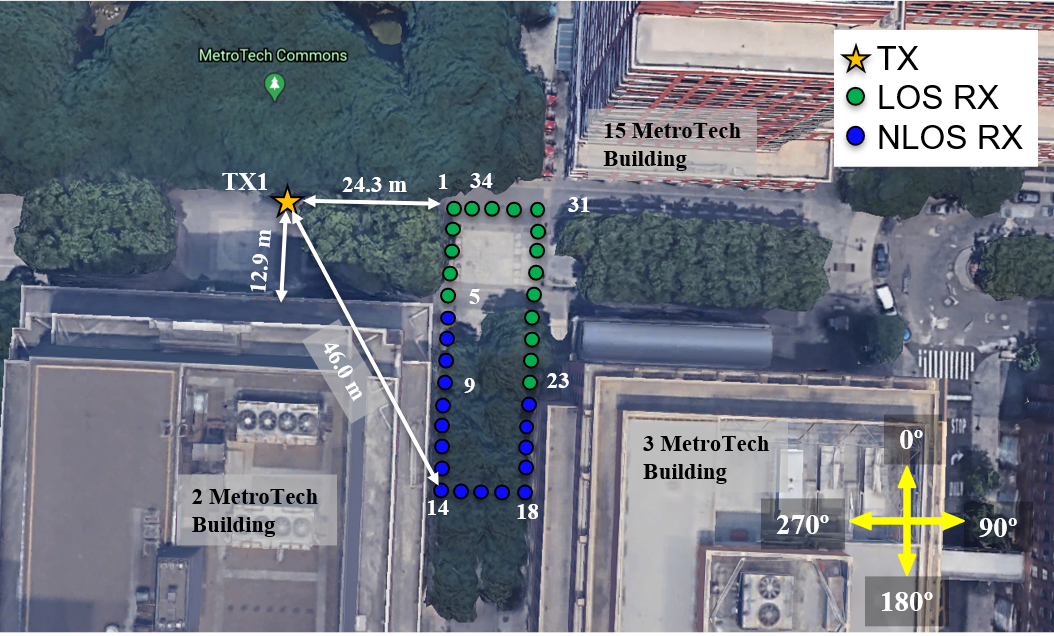}
	\caption{142 GHz outdoor local area measurement locations. }
	\label{fig:map}
\end{figure*}


The 142 GHz local area measurement campaign was conducted in the New York University (NYU) courtyard in Downtown Brooklyn, New York, in 2020. The environment consisted of an open square surrounded by modern buildings covered by infrared-reflective (IRR) glass and concrete walls. Trees and metallic lampposts in the environment caused blockages or reflections. Rotatable high gain narrowbeam horn antennas were used at the transmitter (TX) and receiver (RX) with antenna heights set to 4 m and 1.5 m above the ground to emulate a small-cell base station and a mobile UT, respectively. The TX location was fixed as illustrated in Fig. \ref{fig:map}, and the RX was moved to 34 different locations, which constituted a 39 m $\times$ 12 m rectangular route with a length of 102 m to study the spatial correlation of the received signals. The route shape was motivated by the local area measurements conducted at 73 GHz \cite{Rap17tap}, where a ``L''-shape and a ``C''-shape routes were selected to measure the signal variation when a UT moved around a street corner or in a local area. Locations RX1 - RX5 and RX23 - RX34 (17 locations) were in the LOS scenario; Locations RX6 - RX22 (17 locations) were in the NLOS scenario. Each consecutive and adjacent RX location is 3 m apart from each other. The 2D TX-RX (T-R) separation distance ranged from 24 to 53 m. Such a route design can provide valuable insight on channel spatial variations in path loss, shadow fading, delay spread when the LOS/NLOS transitions occur. Note that the measurements were conducted over several days, and the channel was assumed to be semi-static; thus, the temporal correlation over RX locations was not considered. During the measurements, we observed reflections from three surrounding buildings (MetroTech Buildings 2, 3, and 15) and metallic lampposts. 

A wideband (spread spectrum) sliding correlation-based channel sounder system was used in the 142 GHz measurements, providing a broad dynamic range of measurable path loss of 152 dB \cite{Xing19globecom}. The TX transmitted a wideband pseudorandom noise (PN) signal, and the RX correlated the downconverted received signal with a local copy of the transmitted signal with a slightly offset rate at the baseband to generate a power delay profile (PDP) captured by a high-speed oscilloscope \cite{Mac17sounder}. The channel sounder transmitted a continuous RF signal at center frequency 142 GHz with a 1 GHz radio frequency bandwidth, leading to a minimum time resolution between detectable multipath components equal to 2 ns ($= 1/500$ MHz). Two identical horn antennas with 27 dBi antenna gain and 8\degree~half-power beamwidth (HPBW) were employed at the TX and RX. Two electrically-controlled gimbals mechanically steered the TX and RX antennas with sub-degree accuracy in the azimuth and elevation planes to receive multipath components from all directions. 

At each measurement location, we first searched for the strongest angle of arrival (AOA) and angle of departure (AOD) combination in the azimuth and elevation planes as the starting measurement direction, where the LOS measurements have the TX and RX antennas on boresight and the NLOS measurements have the TX and RX antennas pointing to the best reflection (i.e., strongest signal) direction found by manual search. Then, the TX pointing direction was fixed, and RX swept in the azimuth plane in steps equal to the antenna HPBW (8\degree). 45 stepped-rotations (360/8=45) were performed, and 45 directional PDPs were measured in one azimuth sweep. The RX was then downtilted and uptilted by the antenna HPBW, and we performed the same extensive azimuth sweeps. Overall, three RX sweeps with 135 directional PDPs were recorded for one TX pointing angle. The TX was then pointed to some manually selected directions which have appreciable energy, and the identical three RX azimuthal sweeps were performed for each of the different TX pointing angles. Most RX locations received signals from between one to three TX pointing directions. For each unique TX pointing angle, typically one to three different RX pointing angles are able to provide detectable energy.

\begin{figure}[h!]
	\centering
	\includegraphics[width=1\linewidth]{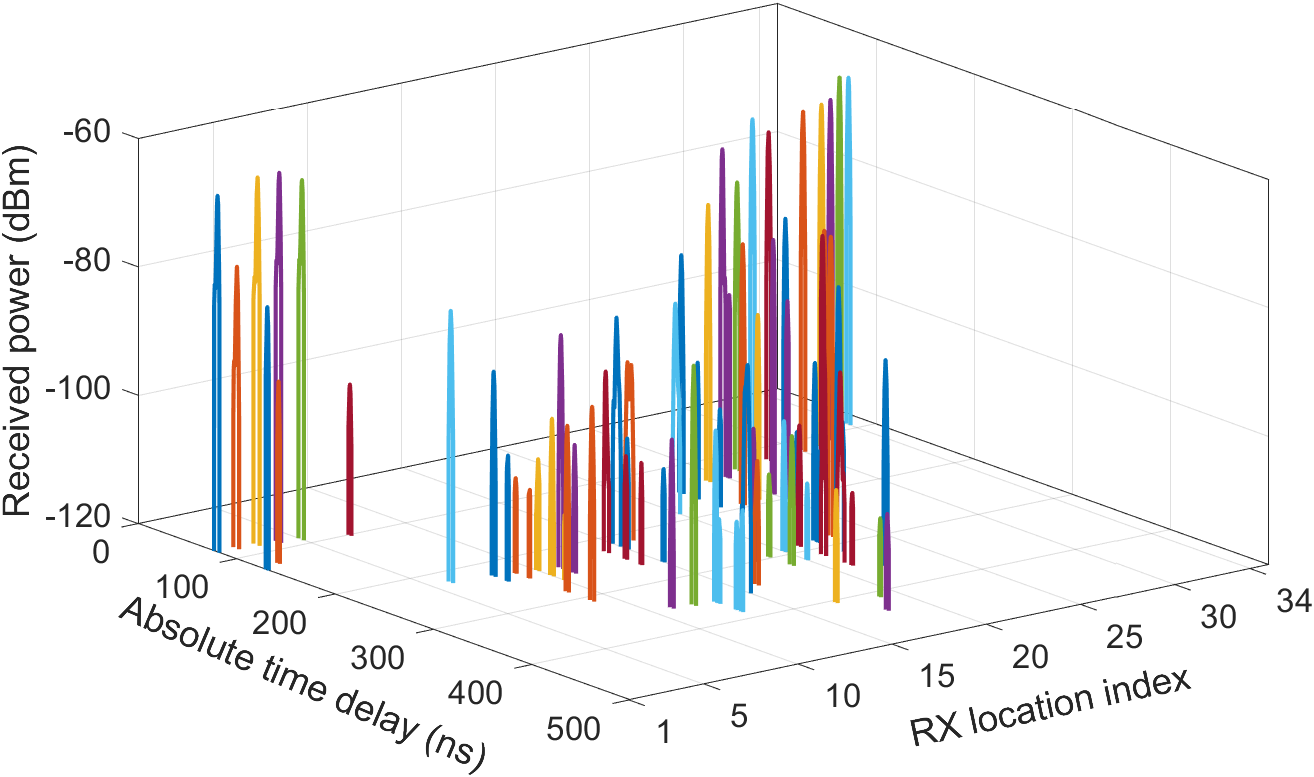}
	\caption{Omnidirectional PDPs with absolute time delays on the links from TX1 to RX1-RX34. }
	\label{fig:omni_pdp}
\end{figure}

Omnidirectional channel characteristics are preferred in channel models since arbitrary antenna patterns can be applied if accurate temporal and spatial statistics are known. To further attain statistics such as delay spread and angular spread of the omnidirectional channels, the measured directional PDPs were synthesized into an omnidirectional PDP by using the absolute timing information provided by a ray tracer NYURay \cite{Kanhere19globecom,Samimi16mtt,Sun15synthesize,Ju21jsac}. The reference clocks at the TX and RX were not connected by a cable and were subject to drift over time. The omnidirectional PDPs with absolute time delays measured at 34 RX locations are shown in Fig. \ref{fig:omni_pdp}.

\section{Large-scale Path Loss along the Rectangular Track}
\begin{figure}
	\centering
	\begin{subfigure}[h]{.5\textwidth}
		\centering
		\includegraphics[width=0.9\linewidth]{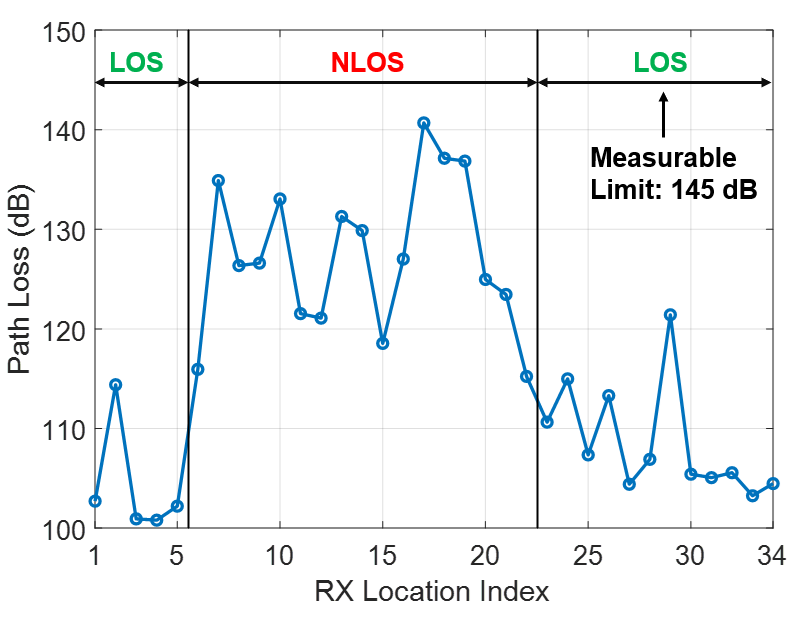}
		\caption{142 GHz omnidirectional path loss variation over 34 RX locations in LOS and NLOS environments.}
		\label{fig:pl_evo}
	\end{subfigure}
	\hfill
	\vspace{.1em}
	\begin{subfigure}[h]{.5\textwidth}
		\centering
		\includegraphics[width=0.9\linewidth]{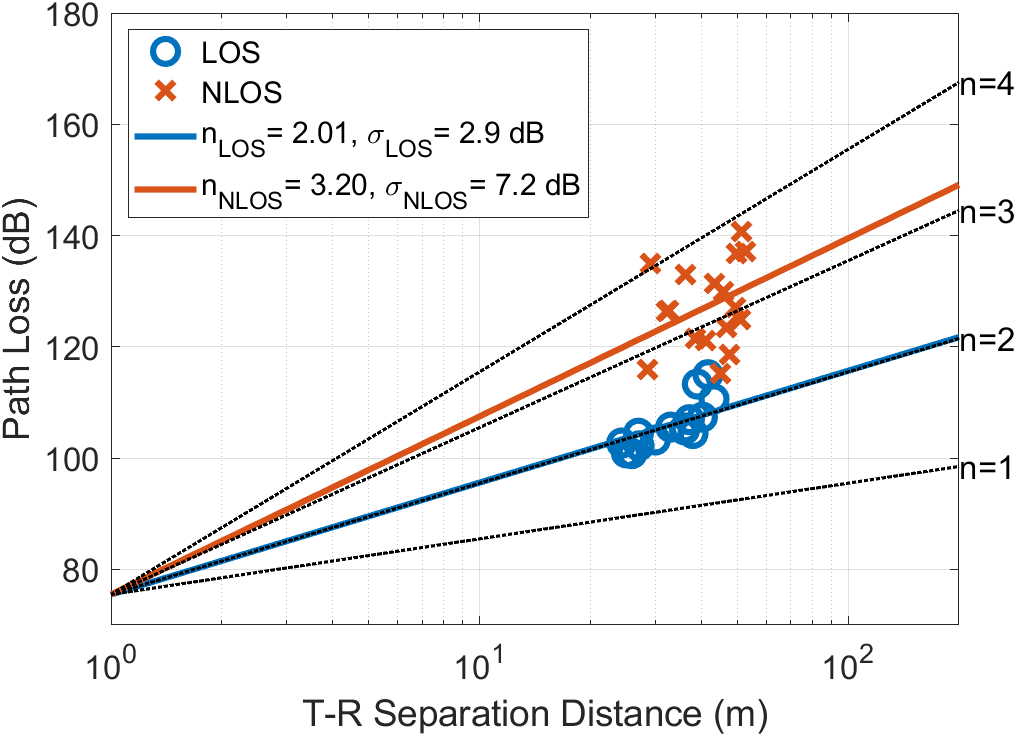}
		\caption{142 GHz omnidirectional path loss scatter plot and CI models for LOS and NLOS measurements.}
		\label{fig:pl_scatter}
	\end{subfigure}
	\caption{Omnidirectional path loss plots for outdoor UMi scenario at 142 GHz.}
	\label{fig:omni_pl}
\end{figure}

Path loss models are used to estimate the signal attenuation over distance and the coverage range for cellular system design. The omnidirectional path loss is calculated by summing all the non-overlapped received powers from directional PDPs measured at one RX location \cite{Sun15synthesize,Samimi16mtt,Ju21jsac}. The omnidirectional path losses measured at 34 RX locations are shown in Fig. \ref{fig:pl_evo}, with LOS and NLOS regions marked in green and red. The path loss increased by 33 dB from RX5 to RX7 (6 m apart) due to LOS-to-NLOS transition, while the path loss decreased by 13 dB from RX21 to RX23 (6 m apart) due to NLOS-to-LOS transition. The increase in path loss from RX5 to RX7 is larger than the decrease from RX21 to RX23 because locations RX5-RX7 were very close to MetroTech Building 2 (the building to the west of these RX locations), which introduced significant shadowing on RX5 to RX7. Work in \cite{Rap17tap} observed that a 25 dB increase in path loss of a LOS-to-NLOS transition measured over similar locations at 73 GHz took a 25 m walk into the urban canyon. However, the path loss increased by 25 dB from RX23 to RX19 at 142 GHz, which were 12 m apart. The larger partition and diffraction loss from the building at 142 GHz causes the faster increase of path loss around the street corner compared to 73 GHz \cite{Xing19globecom,Deng16diffraction}. Such abrupt path loss variations caused by LOS/NLOS transitions should be included in channel models for frequencies above 100 GHz, to realistically model spatial consistency at sub-THz frequencies. 

Note that the LOS measurements at RX2 and RX29 underwent abnormal path loss increases compared to the adjacent RX locations due to the obstruction of several trees in the LOS direction. The total attenuation on the omnidirectional path loss were 10 and 15 dB at RX2 and RX29. However, the measured directional power in the obstructed LOS direction was stronger than the strongest reflection direction by 17 and 4 dB at RX2 and RX29, showing that beam steering and beam selection may not always improve connectivity when the LOS link of a moving UT is temporally blocked by several trees. 


A widely-adopted path loss model for millimeter-wave (mmWave) frequencies, the close-in free space reference distance (CI) path loss model with 1 m reference distance, has been shown to agree with data at sub-THz frequencies \cite{Sun16tvt, Ju21jsac, Xing21a, Xing21b, Xing21icc}:
\begin{equation}
	\label{eq:pathloss}
	\begin{split}
		\textup{PL}^{\textup{CI}}(f,d)[\textup{dB}]=&\textup{FSPL}(f,1\:\textup{m})+10n\log_{10}(d)+\chi_\sigma,
	\end{split}
\end{equation}
where $n$ is path loss exponent (PLE) and $d$ is the T-R separation distance in meters. $\textup{FSPL}(f,1\:\textup{m})$ is the free space path loss at 1 m at the frequency $f$. Shadow fading $\chi_\sigma$ is modeled by a zero mean Gaussian random variable with standard deviation $\sigma$ in dB. PLE $n$ indicates that the power decays by 10$n$ dB per decade of distance beyond 1 m \cite{Rap15tcomm}.


Fig. \ref{fig:pl_scatter} shows the scatter plot of omnidirectional path loss against T-R separation distance at 142 GHz over 32 RX locations (RX2 and RX29 excluded due to the blockage effect). The PLE for the LOS scenario (2.01) is very close to free space (n = 2), and the PLE for the NLOS scenario is 3.20. The PLEs measured at 142 GHz are smaller than the values at 73 GHz (2.53 and 3.61, respectively \cite{Rap17tap}) for both the LOS and NLOS measurements since measurements were conducted at only one TX pointing angle at 73 GHz but one to three TX pointing directions were measured at 142 GHz. Thus, signal powers arriving from other TX pointing angles were not included in the path loss calculation at 73 GHz, and reflections are stronger at higher frequencies \cite{Ju19icc}. Work in \cite{Xing21b} showed that the omnidirectional PLEs measured at 28, 73, and 142 GHz in the indoor environment are very similar, and the additional path loss at higher frequencies may be attributed to the path loss in the first meter of propagation. A small standard deviation of 2.9 dB was observed in the shadow fading for LOS locations, however a larger standard deviation of 7.1 dB was observed for NLOS locations. The rapid birth and death of scattering components bouncing off the surrounding buildings lead to the larger variation in path loss for the NLOS locations. 


\section{Spatial Autocorrelation}
Analysis of spatial autocorrelation of large-scale and small-scale parameters is crucial for wireless system design. The spatial autocorrelation function models the channel correlation property over distance in terms of channel parameters such as shadow fading, delay spread, angular spread. Let $\mathcal{S} = \{\xi_1,\xi_2,...\xi_N\}$ be the non-null set of the channel parameter $\xi$ at $N$ RX location and $d_{ij}$ denote the 2D Euclidean distance between the $i$th and $j$th RX location. For a given distance granularity $\Delta d$, the spatial autocorrelation coefficient $\rho(d)$ of a channel parameter $\xi$ can be estimated over $S$ at a series of discrete distances $d_k$ as \cite{Zhang08gc}
\begin{equation}
		\label{eq:rho_cal}
	\hat{\rho}(d_k) = \frac{\sum_{(i,j)\in\mathcal{I}_k}(\xi_i-\mu_i)(\xi_j-\mu_j)}{\sqrt{\sum_{(i,j)\in\mathcal{I}_k}(\xi_i-\mu_i)^2}\cdot\sqrt{\sum_{(i,j)\in\mathcal{I}_k}(\xi_j-\mu_j)^2}}
\end{equation} 
where $d_k=k\Delta d, k\in \mathbb{Z}$. $\mu_i$ and $\mu_j$ are the sample means of the sets $\{\xi_i:\xi_i\in\mathcal{S};(i,j)\in\mathcal{I}_k\}$ and $\{\xi_j:\xi_j\in\mathcal{S};(i,j)\in\mathcal{I}_k\}$, respectively. Note that each pair of $i$th and $j$th RX locations is only used once. $\mathcal{I}_k$ is given by
\begin{equation}
	\begin{split}
		\mathcal{I}_k=\{(i,j):(d_{ij}-k\Delta d)\in&[-\Delta d/2,+\Delta d/2); \\
		&\xi_i,\xi_j\in\mathcal{S}\}.
	\end{split}
\end{equation}
Here we use 0.05 m ($\approx$25 wavelengths) as the distance granularity $\Delta d$. $\hat{\rho}(d_k)=1$, $-1$, and $0$ implies perfect positive correlation, perfect negative correlation, and no correlation at separation distance $d_k$, respectively.

\subsection{Spatial autocorrelation coefficient of shadow fading}
The spatial correlation of shadow fading has been well studied for sub-6 GHz and mmWave frequencies \cite{Zhang08gc,Gudmundson91a,3GPP38901r16,Algans02jsac}. The shadow fading at two locations are considered independent if these two locations are separated beyond the correlation distance \cite{Ju18gc}. The autocorrelation coefficient $\hat{\rho}(d_k)$ in (\ref{eq:rho_cal}) is empirically computed from the measurement dataset and is commonly fitted by analytical autocorrelation functions $\rho(d)$. Two widely-used autocorrelation functions are the exponential decaying function and exponential decaying sinusoid function \cite{Zhang08gc}. 

\begin{figure}[h!]
	\centering
	\includegraphics[width=0.9\linewidth]{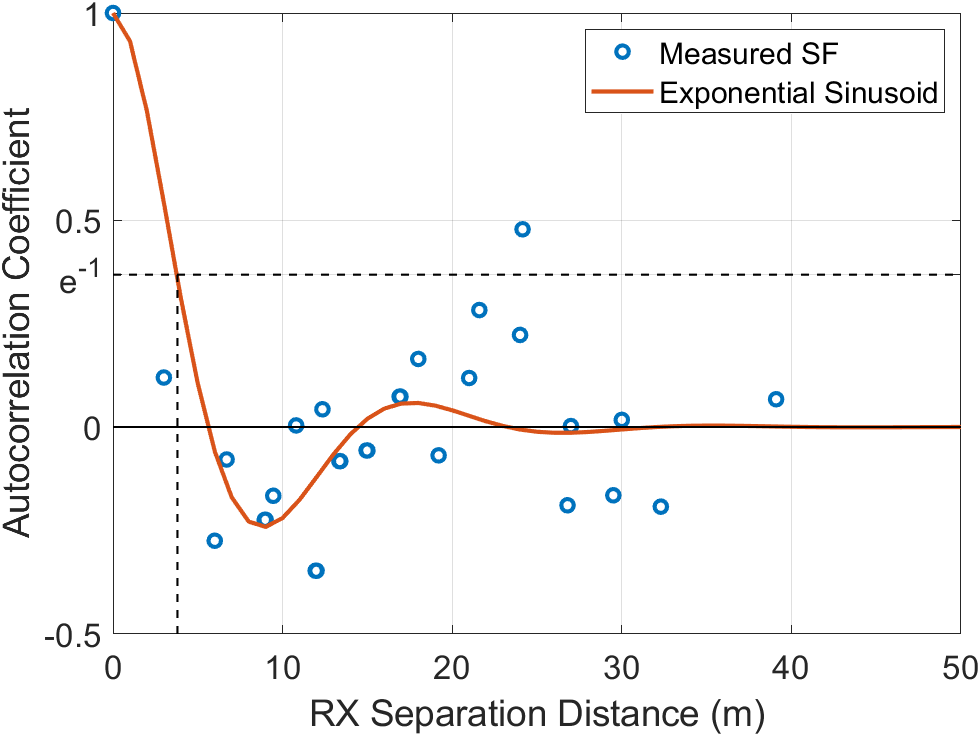}
	\caption{The exponential decaying sinusoid function fit to the spatial autocorrelation function of shadow fading with $D_1$ = 6.2 m and $D_2$ = 2.8 m, with a correlation distance of shadow fading of 3.8 m. }
	\label{fig:sf_corr}
\end{figure}

The exponential decaying function is widely applied in various channel models such as 3GPP and NYUSIM channel models \cite{3GPP38901r16,Ju19gc} to characterize the spatial correlation of shadow fading: $\rho(d) = \exp\left(-d/D_0\right)$. The correlation distance is defined as the distance where the correlation coefficient first drops below 1/e ($\approx 0.37$). If the exponential decaying function is used to model autocorrelation, the correlation distance is equal to $D_0$. For frequencies below 100 GHz, the correlation distance of shadow fading ranges from 10 to 13 m for UMi environment as reported in 3GPP \cite{3GPP38901r16}. However, as shown below, for our particular measurement route the correlation distance is less than 4 m for 142 GHz. The autocorrelation function of shadow fading from the UMi measurements conducted along a rectangular route at 142 GHz does not follow a simple exponential function, as shown in Fig. \ref{fig:sf_corr}, however, the exponential decaying sinusoid function fits the measured correlation data well and is given by \cite{Zhang08gc}:
\begin{equation}
	\label{eq:EDSF}
	\rho(d) = \exp\left(-\frac{d}{D_1}\right)\left[\cos\left(\frac{d}{D_2}\right)+\frac{D_2}{D_1}\sin\left(\frac{d}{D_2}\right)\right],
\end{equation}
The exponential decaying sinusoid function has also been used to model the spatial correlation of received voltage amplitude in small-scale fading measurements \cite{Rap17tap}. Fig. \ref{fig:sf_corr} shows that the exponential sinusoid function with $D_1$ = 6.2 m and $D_2$ = 2.8 m agrees well with the empirical correlation data well at most distances with a root mean square error of 0.18. The correlation distance is 3.8 m. Note that high correlation is observed at a separation distance of 24.15 m which might be caused by the symmetric structure of the street canyon and the rectangular shape of the selected route, due to which RX locations with large separation distances experienced similar shadow fading. Work in \cite{Zhang08gc} showed that the autocorrelation function of shadow fading measured along a single route followed the exponential decaying sinusoid function, even though the autocorrelation function of shadow fading would follow the exponential decaying function when the measured data from multiple routes were lumped together.   

\subsection{Spatial autocorrelation coefficients of delay spread and angular spread}
\begin{figure}[h!]
	\centering
	\includegraphics[width=0.9\linewidth]{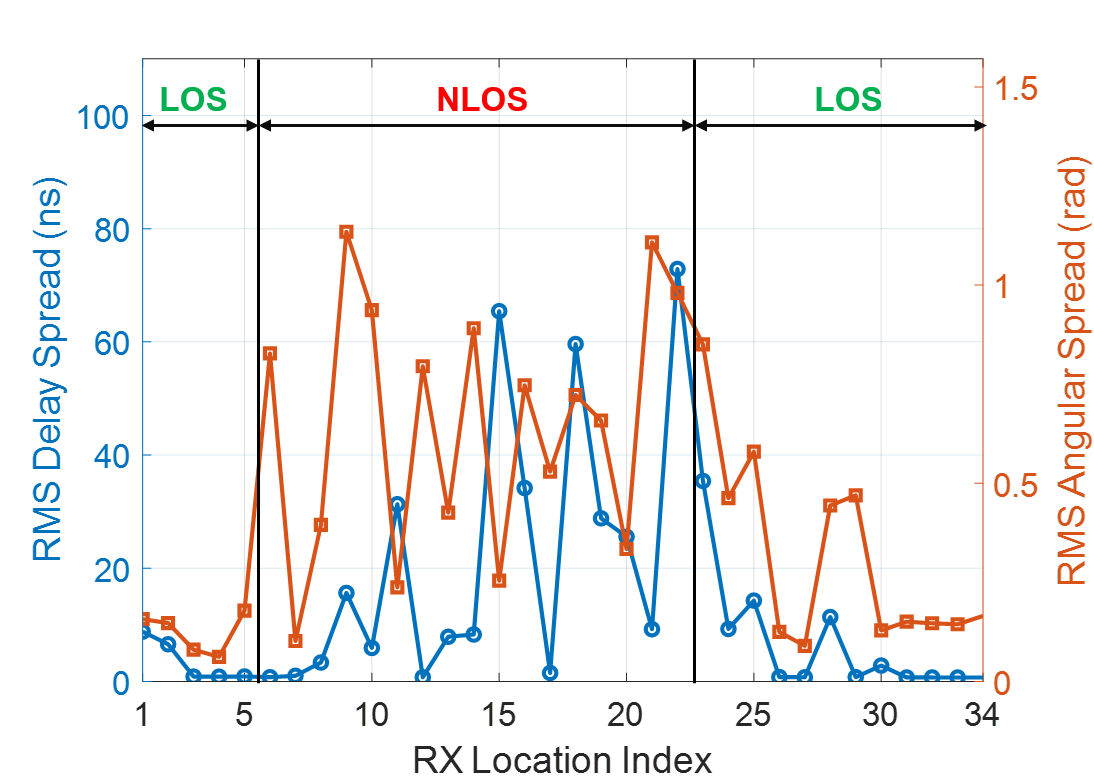}
	\caption{The measured omnidirectional delay spread and angular spread over 34 RX locations in LOS and NLOS environments.}
	\label{fig:ds_as_value}
\end{figure}

Synthesized omnidirectional PDP and omnidirectional power angular spectrum were used to calculate omnidirectional delay spread and AOA angular spread. The RMS delay spread is given by \cite{Rap02textbook}
\begin{equation}
	\sigma(\tau) = \sqrt{\bar{\tau^2}-\bar{\tau}^2},
\end{equation}
where $\bar{\tau}$ is the power-weighted average of multipath component delays, and $\bar{\tau^2}$ is the power weighted average of the square of multipath component delays. Similarly, the angular spread is calculated by \cite{3GPP38901r16,Ju21jsac}
\begin{equation}
	\sigma(\theta)=\sqrt{-2\ln\left(\left|\frac{\sum_{m=1}^M\exp(j\phi_m) P_m}{\sum_{m=1}^{M}P_m}\right|\right)},
\end{equation}
where $\phi_m$ is the AOA or AOD of the $m$th multipath component. The calculated omnidirectional delay spread and AOA angular spread at each RX location are shown in Fig. \ref{fig:ds_as_value}. The mean and standard deviation of delay spread are 5.7 and 8.9 ns for LOS locations and 21.9 and 23.9 ns for NLOS locations; the mean and standard deviation of angular spread are 0.26 and 0.22 radian for LOS locations and 0.65 and 0.32 radian for NLOS locations. LOS RX locations have smaller delay spread and angular spread with low variance due to the dominance of the LOS path; NLOS RX locations generally have larger delay spread and angular spread with high variance, a difference of up to 65 ns in delay spread and 0.9 radian in angular spread when two NLOS RX locations are separated by only 3 m. 

\begin{figure}[h!]
	\centering
	\includegraphics[width=0.9\linewidth]{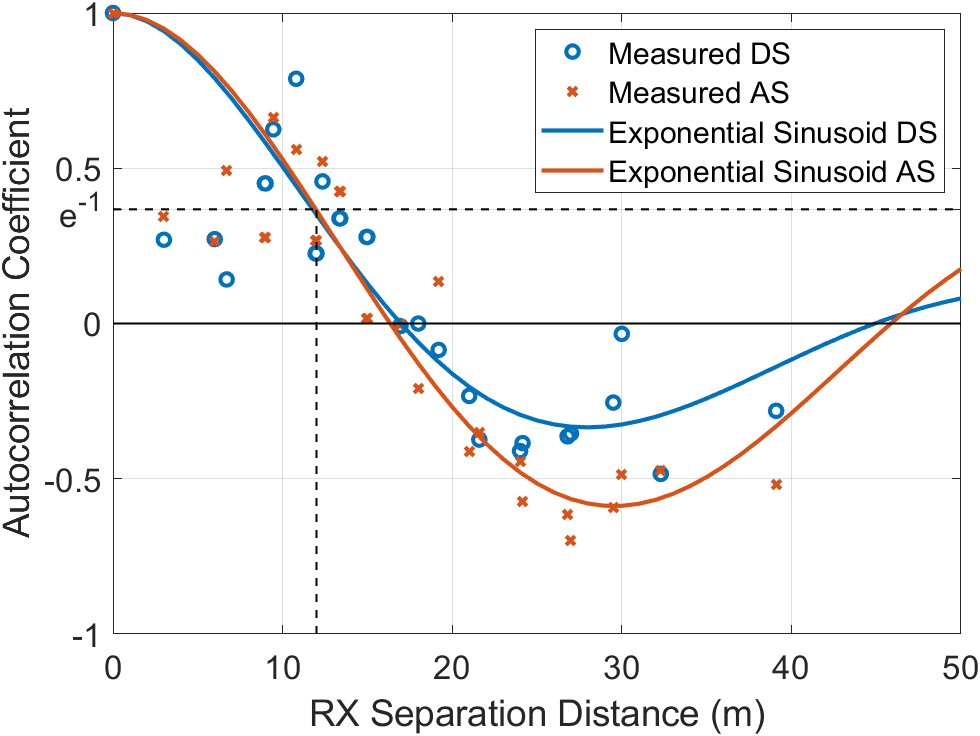}
	\caption{The exponential decaying sinusoid function fits to the spatial autocorrelation functions of delay spread and angular spread where $D1=25.5$ m and $D2=8.9$ m for the delay spread; $D1=55.6$ m and $D2=9.4$ m for the angular spread. The correlation distances for the delay and angular spread are 11.8 m and 12.0 m. DS and AS stand for delay spread and angular spread. }
	\label{fig:ds_as_corr}
\end{figure}
The empirical spatial autocorrelation coefficients of delay spread and angular spread are calculated using (\ref{eq:rho_cal}), as shown in Fig. \ref{fig:ds_as_corr}. We find that the exponential sinusoid function fits the autocorrelation of delay spread and angular spread well. For the delay spread, $D1=25.5$ m and $D2=8.9$ m, while for the angular spread, we find that $D1=55.6$ m and $D2=9.4$ m. The correlation distances for the delay spread and angular spread are 11.8 m and 12.0 m, respectively, indicating that the delay spread and angular spread show a similar decaying trend in the UMi street canyon scenario. Results show that the correlation distances of delay spread and angular spread measured at 142 GHz are close to the values (7-10 m) proposed for frequencies below 100 GHz by 3GPP \cite{3GPP38901r16}. The exponential decaying sinusoid function provides good fits for all three investigated large-scale parameters (i.e., shadow fading, delay spread, and angular spread), suggesting that correlation of channel parameters along a single rectangular route in a local area oscillates. The oscillating correlation may be due to the symmetric structure of the environment and the rectangular route. Therefore, future analysis is required to determine the best correlation model for other asymmetric route shapes in the UMi scenario.

Future work will study the spatial correlation properties of small-scale parameters such as the power and angles of each resolvable multipath component, and the birth and death process of clusters and multipath components. The frequency dependent autocorrelation model will be considered based on the local area measurements at 142 GHz shown in this paper and similar measurements at mmWave frequencies such as 73 GHz \cite{Rap17tap}. Additionally, beam tracking algorithms will be investigated by analyzing the best beam direction at each measurement location.


\section{Conclusion} \label{sec:conclusion}
This paper studied the spatial autocorrelation properties of three critical channel parameters, namely, shadow fading, delay spread, and angular spread, for sub-THz frequencies in a UMi street canyon based on the 142 GHz local area measurements along a 102 m long track. The exponential decaying sinusoid function closely fits the measured autocorrelation functions of shadow fading, delay spread, and angular spread with correlation distances of 3.8 m, 11.8 m, and 12.0 m measured over the rectangular route. The measured correlation distance of shadow fading at 142 GHz of 3.8 m is much shorter than the correlation distance of 10-13 m as specified in 3GPP for frequencies below 100 GHz \cite{3GPP38901r16}. The measured correlation distances of delay spread and angular spread at 142 GHz of 11.8 m and 12 m respectively are comparable to the correlation distances of 7-10 m as specified in 3GPP for frequencies below 100 GHz \cite{3GPP38901r16}. The measurements also indicate that the NLOS locations have more considerable channel variation than the LOS locations. The spatial autocorrelation functions and the corresponding correlation distances presented in this work may be used in spatially consistent channel modeling at frequencies above 100 GHz, to support the design and verification of future advanced beam tracking and beamforming algorithms for 6G and beyond. 

\bibliographystyle{IEEEtran}
\bibliography{icc21}

\end{document}